\documentclass{document-template}

\usepackage[style=numeric, backend=bibtex, sorting=none]{biblatex}
\usepackage[colorlinks=true,linkcolor=blue,citecolor=blue,urlcolor=blue]{hyperref}
\usepackage{url}
\usepackage{subcaption}
\usepackage[utf8]{inputenc}
\usepackage[most]{tcolorbox}
\usepackage{minted}
\usepackage{caption}
\usepackage{adjustbox}

\DeclareUnicodeCharacter{2212}{-}

\bibliography{crc1625_references}

\usepackage{xcolor}

\definecolor{pmdcolor}{HTML}{FFE6CC}
\definecolor{crccolor}{HTML}{F4F5FC}
\definecolor{crccolordark}{HTML}{91BEFF}
\definecolor{crccolordark}{HTML}{91BEFF}
\definecolor{crclightgray}{HTML}{F5F5F5}

\tcbset{
    pmd/.style={
      on line,
      boxrule=0pt, 
      boxsep=0pt, 
      top=2pt,
      left=2pt, 
      bottom=2pt,
      right=1pt,
      colback=pmdcolor,
      colframe=black,
      coltext=blue!10!black,
      fontupper={\ttfamily\small\strut}
    },
    crc/.style={
      on line,
      boxrule=0pt, 
      boxsep=0pt, 
      top=2pt,
      left=2pt, 
      bottom=2pt,
      right=1pt,
      colback=crccolor,
      colframe=black,
      coltext=blue!80!black,
      fontupper={\ttfamily\small\strut}
    },
    entity/.style={
      on line,
      boxrule=0pt, 
      boxsep=0pt, 
      top=2pt,
      left=2pt, 
      bottom=2pt,
      right=1pt,
      colback=crclightgray,
      colframe=black,
      coltext=blue!10!black,
      fontupper={\ttfamily\small\strut}
    },
}

\newtcolorbox{bluebox}[1]{
  colback=crccolordark!5!white,
  colframe=crccolordark!75!black,
}

\begin{document}

\pagestyle{fancy}

\title{Field report from Collaborative Research Center 1625:\\
\mbox{Heterogeneous} research data management using ontology\\ \mbox{representations}}

\maketitle

\author{Doaa Mohamed}
\author{Samuel García Vázquez}
\author{Behnam Mardani}
\author{Victor Dudarev}
\author{Alfred Ludwig}
\author{Maribel Acosta}
\author{Markus Stricker*}

\begin{affiliations}
D. M., B. M., M. S.\\
Interdisciplinary Centre for Advanced Materials Simulation, Ruhr-Universit\"at Bochum, Universit\"atsstra\ss e 150, 44801 Bochum, Germany\\
Email Address: $^*$markus.stricker@rub.de\\

V. D., A. L.\\
Materials Discovery and Interfaces, Ruhr-Universit\"at Bochum, Universit\"atsstra\ss e 150, 44801 Bochum, Germany\\

S. G. V., M. A.\\
Data Engineering Research Group, Technical University of Munich, Bildungscampus 2, 74076 Heilbronn, Germany\\
\end{affiliations}

\keywords{research data management, materials science, ontology, knowledge graph, electrocatalysis, materials characterization, machine learning}

\begin{abstract}
The goal of the Collaborative Research Center 1625 is the establishment of a scientific basis for the atomic-scale understanding and design of multifunctional compositionally complex solid solution surfaces.
Next to materials synthesis in form of thin-film materials libraries, various materials characterization and simulations techniques are used to explore the materials data space of the problem.
Machine learning and artificial intelligence techniques guide its exploration and navigation.
The effective use of the combined heterogeneous data requires more than just a simple research data management plan.
Consequently, our research data management system maps different data modalities in different formats and resolutions from different labs to the correct spatial locations on physical samples.
Besides a graphical user interface, the system can also be accessed through an application programming interface for reproducible data-driven workflows.
It is implemented by a combination of a custom research data management system designed around a relational database, an ontology which builds upon materials science-specific ontologies, and the construction of a Knowledge Graph.
Along with the technical solutions of research data management system and \textit{lessons learned}, first use cases are shown which were not possible (or at least much harder to achieve) without it.
\end{abstract}

\section{CRC1625 objectives and structure}
This field report presents the collaborative research data management and heterogeneous data usage efforts within the Collaborative Research Center 1625 (CRC1625) ``Atomic-scale understanding and design of multifunctional compositionally complex solid solution surfaces''.
The CRC1625 aims to use the potential of compositionally complex solid solutions (CCSS) as versatile materials design platforms for future electrocatalysts by developing a comprehensive theoretical and experimental understanding of their atomic-scale surface features~\cite{Batchelor2021}.
The unique properties of CCSS come from the many different poly-elemental active sites on their surfaces.
By gaining control and design capabilities over these surface atom arrangements (SAAs), we aim to improve upon the limitations of conventional electrocatalysts and create multifunctional materials with new combinations of activity, stability, and potential for cascade reactions.

SAAs, which are specific arrangements of (sub)surface atoms and their chemical identities, are found in large numbers and varieties on CCSS surfaces.
Understanding SAAs is important for linking composition, structure, and activity for CCSS surfaces.
We enable the application of data-driven approaches through our research data management.
Our data-driven workflows combine data from different scales and research fields, extract knowledge from data, and organize it within Knowledge Graphs.
This comprehensive approach supports the CRC’s goal to control SAAs at the atomic scale across surfaces, allowing us to design ideal CCSS surfaces for specific electrochemical applications.

The CRC1625 involves 19 individual projects at five different institutions.
Most of these projects produce heterogeneous research data from simulations, synthesis, baseline characterization, atomic-scale characterization, and electrochemical characterization.
Three projects are solely at the receiving end of the data produced in all others:
One project provides the infrastructure and support for research data management~\cite{Dudarev2025}; the second of these three is tasked with the establishment of a Knowledge Graph for the design of CCSS surfaces; the third focuses on data-guided experimentation using machine learning.
Our consortium structure is a design choice tailored to the general problem space of materials: Neither the fastest high-throughput experimental screening methods nor any high-throughput simulation approaches can (in reasonable time) cover the required composition-property space at the relevant resolutions to brute-force establish the scientific basis for CCSS as a discovery platform for multifunctional, tunable materials.
Namely, there is a need for data-based methods which can speed up characterization procedures through active learning~\cite{Stricker2022e,Thelen2023,Thelen2025b} or in the prediction of candidates for reaction-specific high-performing compositions~\cite{Zhang2025a,Zhang2025d,Zhang2025c} where our data serves in training and validation.
All these data-based approaches naturally require FAIR data~\cite{wilkinson2016fair} and code~\cite{Barker2022}.

In this report we outline the materials space in which the CRC1625 is situated, i.e., the types of characterization techniques and computational methods used and the challenges associated with these individual data modalities in their raw and processed form, as well as the challenge to combine all these data sources.
In addition, we show highlights of how our research data management system (RDMS) supports our data-driven workflows, and how the Knowledge Graph builds a semantic layer upon it.
It is our hope that the community finds this report useful, and we look forward to constructive, inspiring discussions.

To help the reader appreciate our data scope, we explicitly outline all the data modalities we produce and manage in the next two subsections.
The majority of data originates from experimental characterization w.r.t. composition, structure, and electrochemical characterization.
Data of experimental origin are supplemented with density-functional theory (DFT) simulations, molecular dynamics simulations, and reused in machine learning approaches.

\subsection{Experimental methods}
\begin{figure}[htp!]
\centering
\begin{subfigure}{.49\textwidth}
    \includegraphics[width=\textwidth]{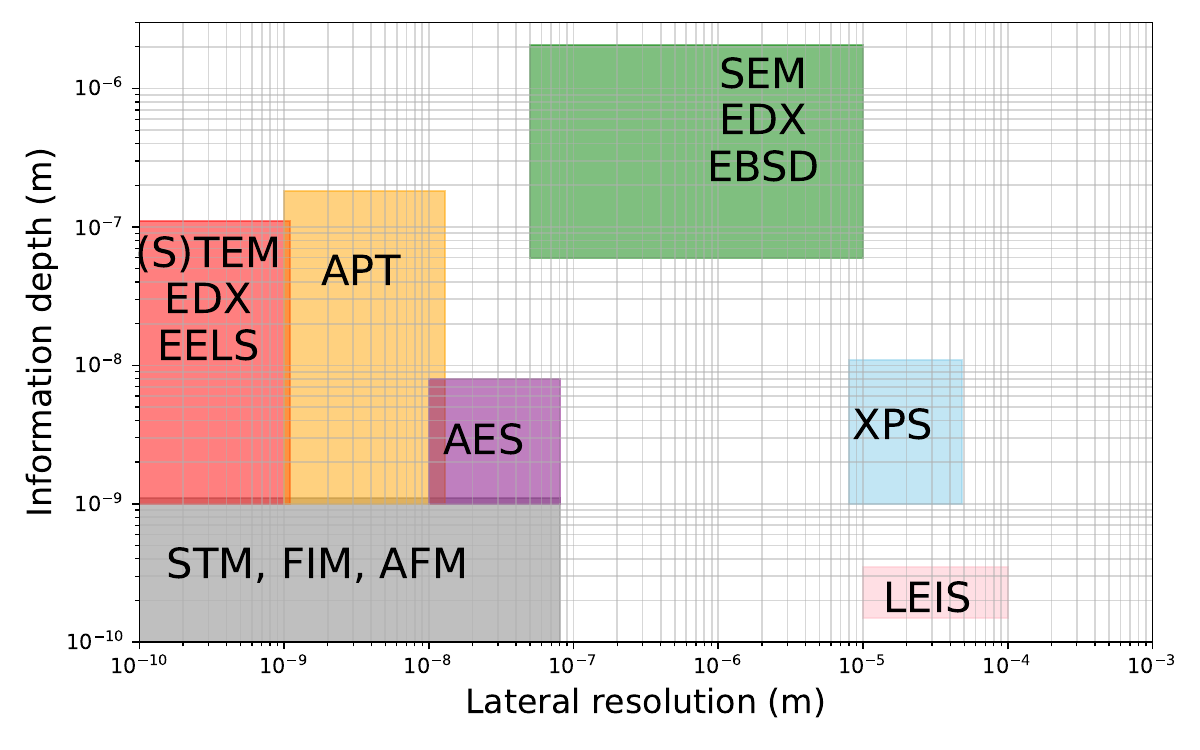}
    \caption{Chemical and structural characterization.}
    \label{fig:chem_char_res}
\end{subfigure}
\hfill
\begin{subfigure}{.49\textwidth}
    \includegraphics[width=\textwidth]{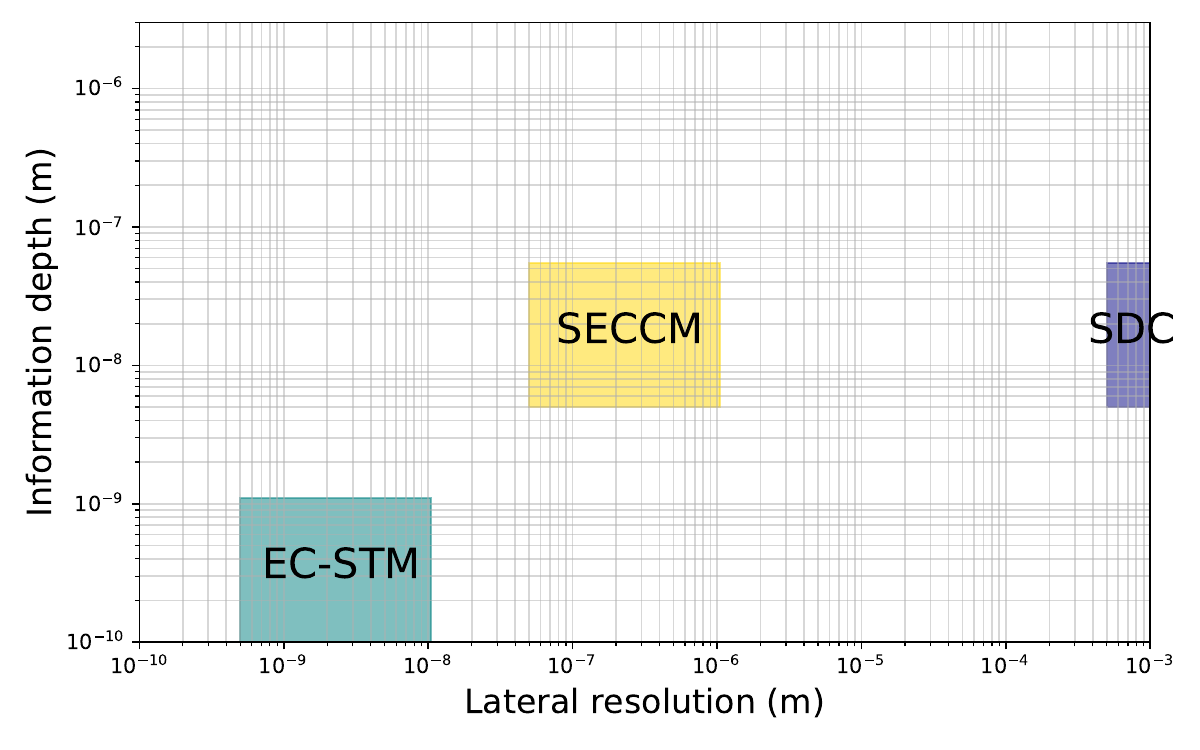}
    \caption{Electrochemical characterization.}
    \label{fig:electrochem_char_res}
\end{subfigure}
\caption{Approximate resolution ranges of compositional and/or structural Figure~\ref{fig:chem_char_res} and electrochemical Figure~\ref{fig:electrochem_char_res} characterization techniques for surfaces employed in CRC1625. Only the compositional/structural techniques also probe to some appreciable level of depth (Y-axis). The electrochemical characterization technique's depth is estimated. Refer to Table~\ref{tab:char_techniques} for the explanation of the abbreviations.}
\label{fig:char_techniques}
\end{figure}

Our main sample preparation method is combinatorial magnetron co-sputtering of multiple elemental sources to obtain thin-film materials libraries covering large compositional ranges.
Based on scientific considerations and data-guided exploration, e.g. from high-throughput simulations, promising compositions ranges are identified for synthesis and then characterized in high-throughput mode.
The key challenge in CRC1625 is the atomic-scale characterization of CCSS and being able to determine the SAA with a focus on chemical identification which goes beyond the determination of average surface compositions.
We neglect structural characterization because we stay in FCC solid solutions and up to now do not consider reconstructions.
We address the challenge of identifying individual SAA with a variety of methods with different lateral resolutions and information depths, cf. Figure~\ref{fig:chem_char_res}, from near atomic resolution to more integrated measures.
The methods also vary w.r.t. their information depth, i.e., their surface sensitivity.
Energy-dispersive X-ray spectroscopy (EDX), for example, allows us to assess volumetric elemental concentrations.
In addition to the chemical and structural characterization we also employ three electrochemical characterization techniques from the near atomic scale electrochemical scanning tunneling microscopy (EC-STM), high-resolution (nanoscale) scanning electrochemical cell microscopy (SECCM), and fast screening through scanning droplet cell microscopy (SDC), cf. Figure~\ref{fig:electrochem_char_res}.

These characterization techniques provide us with a scale-spanning, multimodal view of the chemical, structural, and electrochemical properties of the materials libraries and pose a challenge for research data management, mainly in the connection of information at different scales.
Not all techniques are applied on all samples, and particularly the high-throughput techniques produce much more data than high-resolution techniques, which naturally creates an imbalance and, therefore, sparsity in the data records for specific compositions.
Table~\ref{tab:char_techniques} provides an overview of the different data types associated with each technique and the current data volume per technique.
This overview shows that our main challenge in managing our data is not the volume but lies the complex relationships between individual data points.

\begin{table}[htp!]
\caption{Overview of structural and chemical characterization techniques, number of different data formats, and their respective data volume as of October 2025. The total volume of the data in the RDMS currently is 11.3 GB.}
    \centering
    \begin{tabular}{|l|l|r|r|}
    \hline
    Abbreviation & Description & \# different data formats & Current data volume (MB)\\
    \hline \hline
    STM & Scanning tunneling microscopy & 1 & 5\\
    FIM & Field ion microscopy & 1 & 3989\\
    AFM & Atomic force microscopy & 5 & 32\\
    (S)TEM & (Scanning) transmission electron microscopy & 4 & 2332\\
    EDX & Energy-dispersive X-ray spectroscopy & 26 & 634 \\
    EELS & Electron-energy loss spectroscopy & n/a & n/a\\
    APT & Atom probe tomography & 2 & 596 \\
    AES & Auger electron spectroscopy & n/a & n/a\\
    SEM & Scanning electron microscopy & 6 & 311\\
    EBSD & Electron backscatter diffraction & 1 & 322\\
    XPS & X-ray photoelectron spectroscopy & 5 & 21 \\
    LEIS & Low-energy ion scattering & 0 & 0\\
    \hline\hline
    EC-STM & Electrochemical scanning tunneling microscopy & n/a & n/a\\
    SECCM & Scanning electrochemical cell microscopy & 4 & 3042\\
    SDC & Scanning droplet cell microscopy &4 & 63\\
    \hline
    \end{tabular}
    \label{tab:char_techniques}
\end{table}

\subsection{Computational methods}
The computational methods employed in CRC1625 encompass mainly atomistic techniques: density functional theory (DFT) and molecular dynamics (MD) using machine-learned interatomic potentials based on the graph atomic cluster expansion (GRACE) formalism~\cite{Lysogorskiy2025}.
These GRACE-based approaches are used to predict segregation and ordering at CCSS surfaces; DFT is used to understand atomic-scale processes of electrochemical reactions at CCSS surfaces and as training data for fitting surrogate models for the prediction of the adsorption energies of specific SAA~\cite{Batchelor2021,Clausen2023}.
In addition to these \textit{classical} simulation techniques, we employ a number of data-driven techniques based on \textit{all} of the available data to accelerate characterization and to learn design principles of CCSS surface for specific electrocatalytic applications.

In this article, we focus on the experimental data modalities from now on and provide examples for data-driven workflows based on experimental data in Section~\ref{sec:ai_readiness}.
This has two reasons: 1) Data management for atomistic simulation output is largely solved in the community, e.g., through the OPTIMADE API~\cite{Evans2024}; and 2) the number of projects in CRC1625 employing experimental techniques is much larger and, consequently, the heterogeneity of the raw data is much higher and more challenging to integrate.

\section{Research data management system (RDMS)}
The core challenge of research data management for such a heterogeneous data space is an appropriate coordinate system to navigate this high-dimensional data space.
We approach this challenge on several levels: on the level of data (RDMS), on the level of data modeling (ontology, cf. Section~\ref{sec:ontology}), and on the level of data representations for machine learning (AI-readiness, cf. Section~\ref{sec:ai_readiness}).
We begin with outlining our solutions for different raw data formats and provide infrastructural details of our RDMS.

\subsection{System design and architecture}
In recent years, materials science has witnessed a noticeable shift due to the advancements in information technology, data science, and automation, which are now generating more data than ever -- from high-throughput experiments to large-scale simulations.
But this growth is accompanied by challenges for the management of the created data.
Datasets are often large (where \textit{large} here should be seen relative to the effort to obtain the data in the first place), diverse, fast-changing, and sometimes difficult to trust or use effectively, e.g., because of missing metadata or undocumented or even unknown uncertainty.
In particular, the heterogeneous nature of the materials data space poses a significant challenge to integrating data from different sources.
These challenges, often described with the 5 Vs of big data -- volume, variety, velocity, veracity, and value -- make it clear that a good RDMS is no longer optional; it is essential~\cite{medina2022accelerating}.
An ideal RDMS needs to follow the FAIR principles: the data should be Findable, Accessible, Interoperable, and Reusable.
Most researchers agree that good data management is important, but in practice the development and adoption of standards for data structure, even taxonomy, is still a work in progress, e.g., through national consortia like the NFDI MatWerk~\cite{nfdimatwerk2024}.
As a result, many labs still manage their data on their own with custom systems which are usually incompatible with other labs and/or the wider community.
In CRC1625, we use MatInf~\cite{Dudarev2025}, an open-source, internet-accessible platform designed to support research in materials science.
It is flexible and simple to adapt to different labs and their data with associated metadata.
With support for representing complex materials systems, different property types, and a user-friendly interface inspired by the periodic table, MatInf helps researchers organize, track, and explore their data~\cite{Dudarev2025}. 

\subsection{Data structure and functionalities}
MatInf is specifically designed to handle data format diversity by supporting various high-level categories of data:

\begin{itemize}
    \item Experimental data: includes both raw and processed measurements obtained from techniques such as high-throughput experimentation, electron microscopy (e.g., SEM and EDX), spectroscopy, and electrochemical characterization. Data may come in various formats, including CSV files for numerical data, image formats such as JPEG and TIFF, and compressed ZIP archives containing bundled datasets -- all of which are supported by the system.
    \item Metadata: we capture rich metadata that describes the context of experiments, such as sample identifiers, synthesis details, instrument settings, and user actions. This metadata is the basis for reusability and traceability.
    \item Relational and hierarchical links: allow users to connect samples, analysis steps, and results in a meaningful way. This technique makes it easy to trace how data was generated and how it fits into the broader research process (workflow).
    \item User-generated content: such as hypotheses, ideas, comments, workflow definitions, and annotations, is treated as an integral part of the data space. This helps improve collaboration and reproducibility across research teams and track the provenance of ideas and the rationale \textit{why} a certain composition was synthesized and who is to be credited (authorship) even if the resp. researcher is no longer associated with the project.
\end{itemize}

\begin{figure}[htp!]
    \centering
    \includegraphics[width=.75\textwidth]{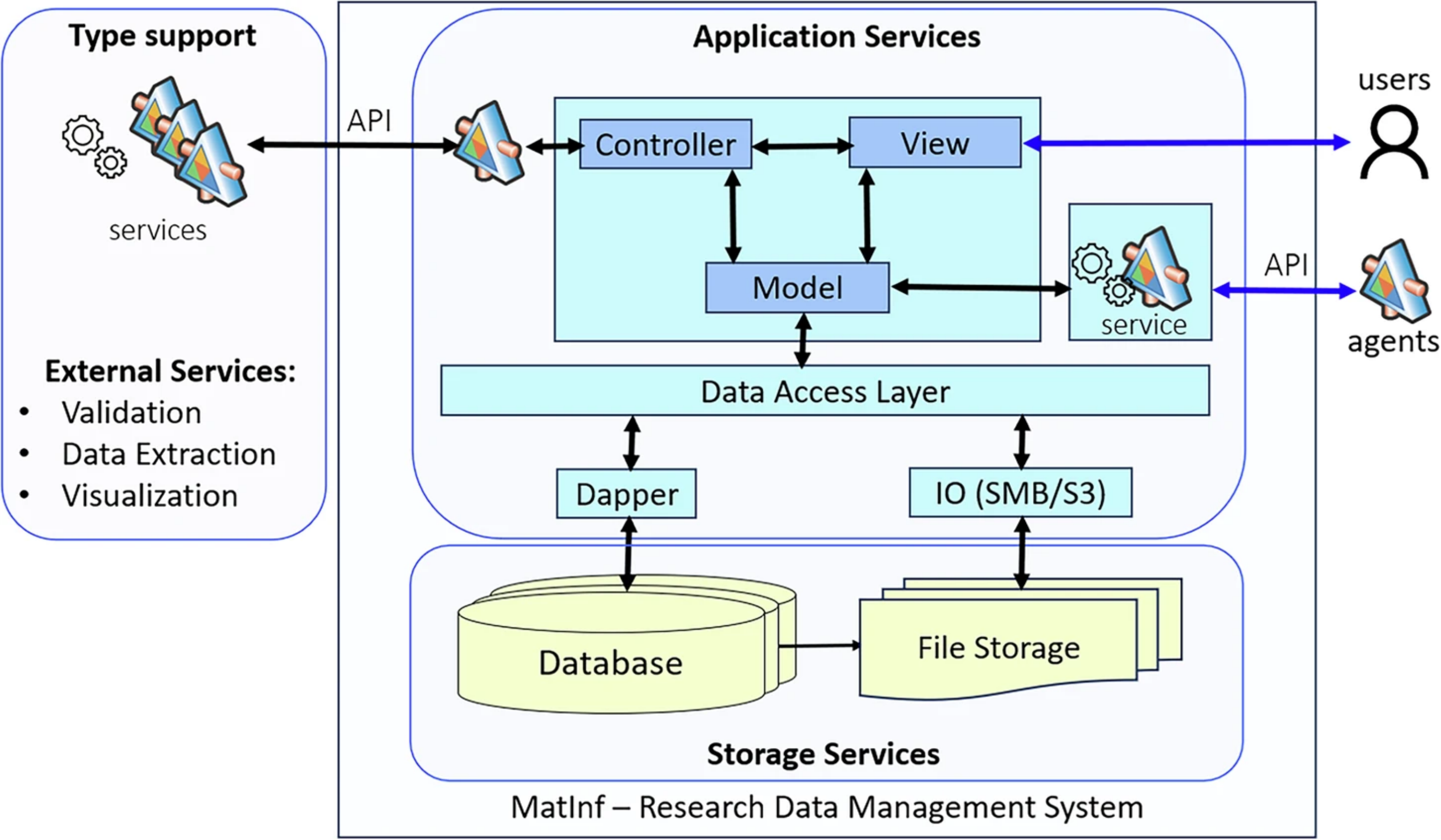}
    \caption{Schematic of the architecture of the MatInf Research Data Management System. The system follows a modular three-tier design, comprising application, data access through an API, and storage layers. It integrates RESTful APIs for communication with users and agents, ensuring extensibility through external services such as validation, data extraction, and visualization.
    From~\cite{Dudarev2025}, licensed under \href{http://creativecommons.org/licenses/by/4.0/}{Creative Commons Attribution 4.0}.}
    \label{fig:MatInf}
\end{figure}

The backend of MatInf is built on a modular three-tier architecture, combining an ASP.NET Core application layer with a relational Structured Query Language (SQL) server database.
This design provides reliability and scalability while maintaining data integrity through well-structured relational and customizable schemas.
Core scientific entities such as materials systems, compositions, and modifications are stored in normalized tables, while additional attributes are managed through type-specific property tables.
This provides a robust infrastructure for data and remains flexible enough to support new or evolving information.
The modular design of the MatInf backend (Figure~\ref{fig:MatInf}) highlights the separation between application logic, data storage, and external service integration, supporting scalability and FAIR-compliant interoperability.

Interoperability is achieved through RESTful APIs that follow the OpenAPI specification.
These interfaces cover standard operations for creating, reading, updating, and deleting objects and metadata but also extend to more advanced services such as file validation, structured data extraction, and visualization.
A typical workflow allows files to be uploaded, checked for consistency, automatically parsed into structured JSON or CSV format, and directly visualized in the browser or through external services connected via the API.
In particular, the User-Defined Type (UDT) Support API enables users to define custom data types and integrate new formats into the system. Through this interface, external web services can handle validation and automatic data processing according to configurable type settings, including visualization of uploaded files. This mechanism allows the RDMS to flexibly adapt to evolving research needs and ensures interoperability across diverse data workflows.
For detailed information about creating and configuring UDTs, see the \textit{UDT Support API} section in the documentation (\url{https://l.rub.de/dc738632}).
The API layer also allows connection with external repositories, measurement instruments, and machine learning pipelines, turning the system into both a laboratory-level tool and a backend for larger, federated infrastructures.

Metadata forms the backbone of our system, ensuring that all research objects (samples, measurements, datasets, etc.) are findable, interpretable, and reusable.
Each object carries contextual information such as unique identifiers, measurement configurations, contributor actions, and timestamps.
By linking physical with virtual objects, the link between measurements performed on samples and their associated data files maintains a record of provenance that allows traceability and enables reproducibility across laboratories and projects.
The structure of the metadata layer is flexible and allows individual data creators to extend descriptions when new experimental methods or properties are introduced.
In this way, metadata captures not only the identity of research objects but also their connection with other research objects within a workflow, making it a central enabler of FAIR principles in our research environment.

\begin{figure}[htp!]
    \centering
    \includegraphics[width=.9\textwidth]{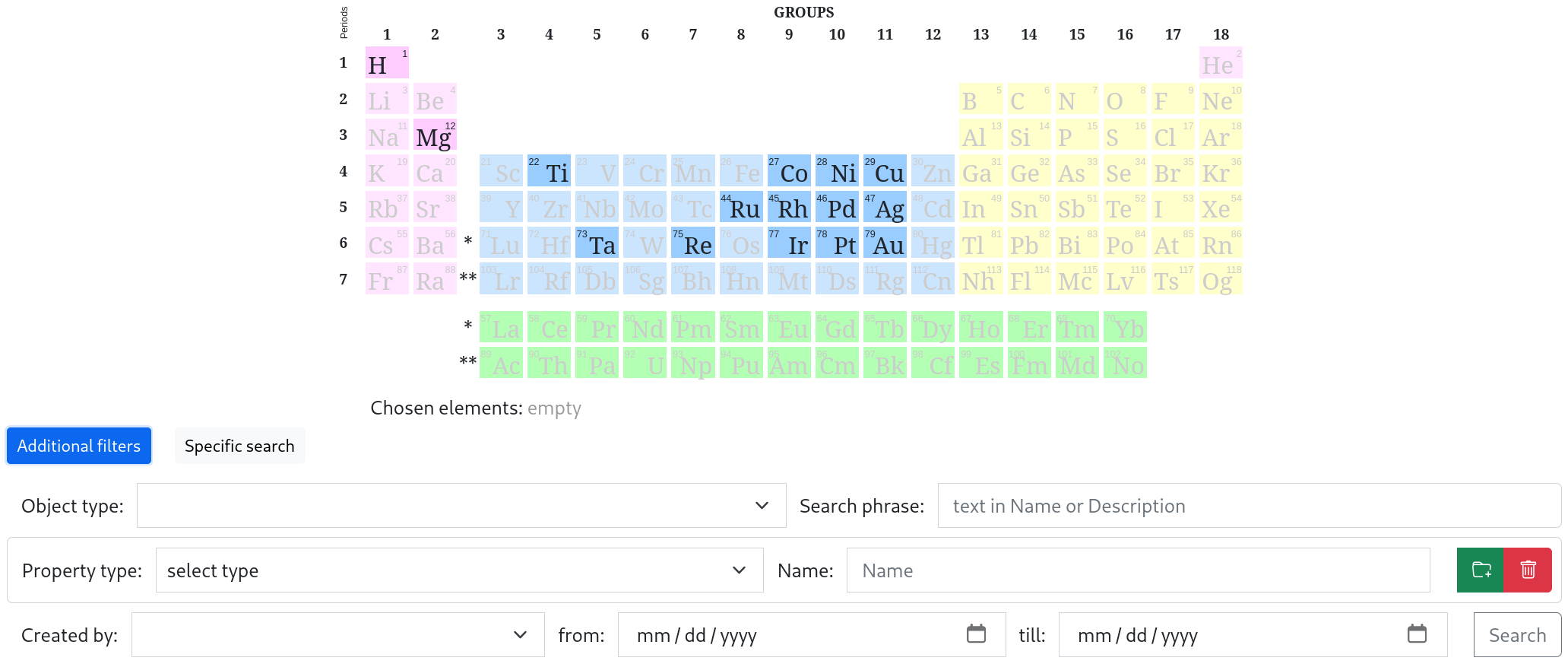}
    \caption{Screenshot of graphical user interface's search function with a periodic table for chemical system and additional filtering options for object type, search phrase, property type, name, creator (author), and time range.}
    \label{fig:matinf_gui_periodic_table}
\end{figure}

\begin{figure}[htp!]
    \centering
    \includegraphics[width=1\textwidth]{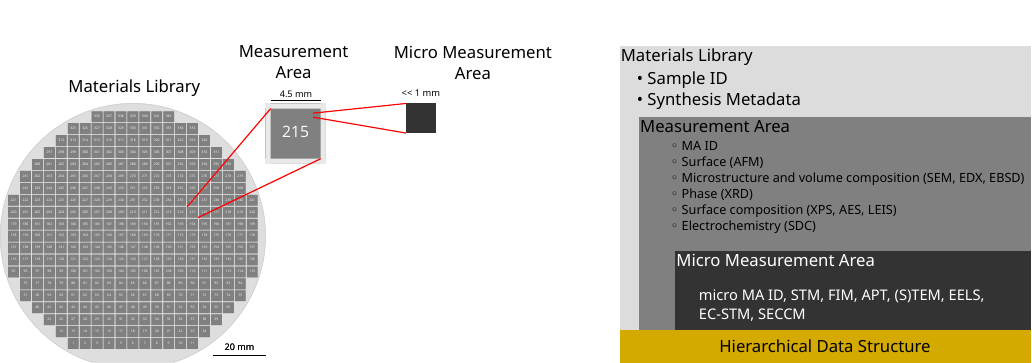}
    \caption{Left: Real space origin of characterization location on materials library (ML), measurement area (MA), micro measurement area (MMA); right: hierarchical, multidimensional and multimodal data structure.}
    \label{fig:wafer_data_dimensions}
\end{figure}

The graphical user interface (GUI) is designed to provide intuitive access to complex datasets without requiring technical expertise.
Figure~\ref{fig:matinf_gui_periodic_table} shows a screenshot of our main search function: Users can click on elements to define the chemical system (elements contained in the measurement area (MA) of a sample, cf. Figure~\ref{fig:wafer_data_dimensions}).
Non-existing elements in the RDMS are shown with a fainter color.
Additional filters for object type, search phrase, property type, name, creator, and a time span allow granular search.
The GUI can also be used to upload files, annotate datasets, browse object relationships, and visualize experimental results as well as validation and preview functions.
This enables contributors to check data quality w.r.t. format and content immediately after upload.
Interactive components support the exploration of hierarchical relationships between samples, measurements, and analyses, helping users understand dependencies and provenance.

The underlying relational concept strictly follows the physical objects' spatial relationships.
Figure~\ref{fig:wafer_data_dimensions} shows the real space connection (left) with our hierarchical data formats (right) for all characterization techniques (cf. Figure~\ref{fig:char_techniques}):
Each smaller-scale characterization (from materials library `sample' to MA to micro measurement area (MMA)) inherits all properties from the larger scale, e.g., Sample ID;
each materials library is discretized with a 342-grid of MAs spanning 4.5\,mm by 4.5\,mm. Each MA can be characterized with higher-resolution techniques $\ll 1\,$mm which we denote micro measurement area (MMA).
Depending on the technique, this can be at or near atomic resolution.

While it is generally possible to fill all \textit{dimensions} of a sample, MA, and MMA, we typically have sparse datasets of materials libraries.
This sparsity is prevalent in materials science: many materials/compositions are only characterized w.r.t. certain properties which are relevant for the research question.
Additional measurements are usually based on a cost-benefit decision.
To some extent, this sparsity can be addressed with machine learning techniques (e.g. inferring correlations or via interpolation, cf. Section~\ref{sec:ai_readiness}) but even partial coverage allows for efficient exploration of the compositions-property space in our case.

\begin{figure}[htp!]
\centering
\includegraphics[width=0.8\textwidth]{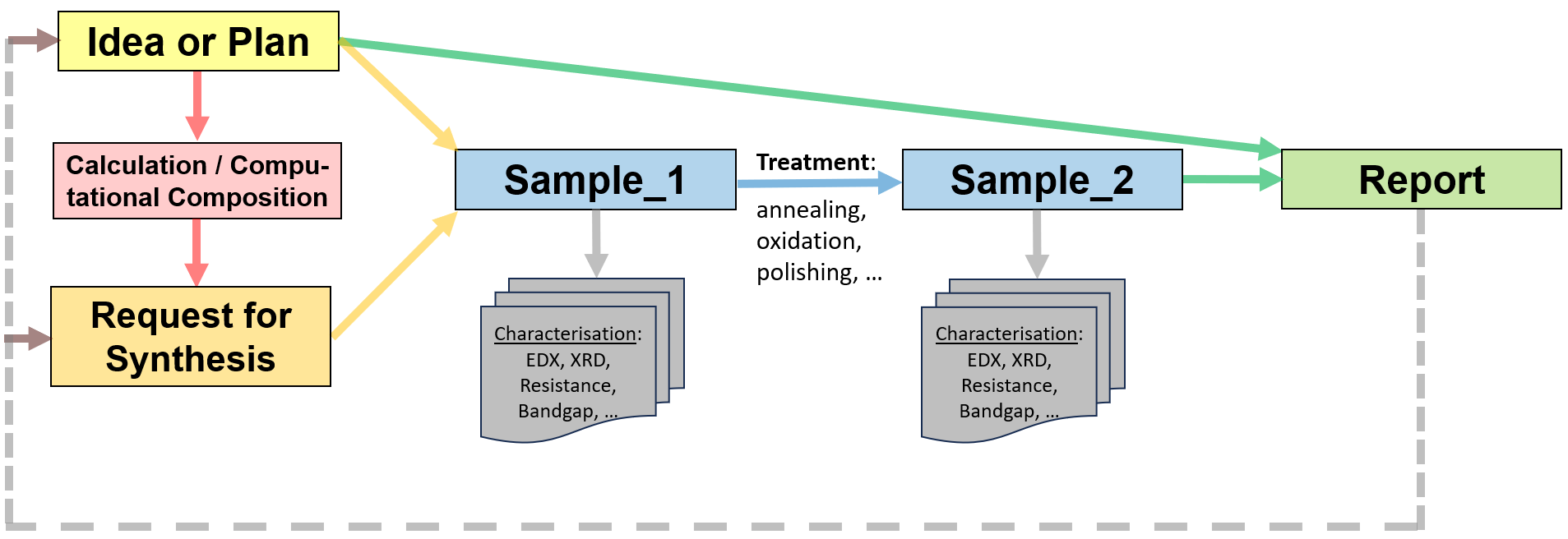}
\caption{Sample-centric workflow within the CRC1625 RDMS, showing relationships between planning, synthesis, characterization, and reporting \cite{Dudarev2025a}. 
}
\label{fig:workflow}
\end{figure}

In our sample-centric hierarchy, we can extract the synthesis and characterization history as a workflow (see Figure~\ref{fig:workflow} and~\ref{fig:ontology_overview}).
In addition, it is also possible to explicitly define a workflow a sample has to go through a `handover object', which models that a physical sample has to change location to different groups within the CRC1625 for different characterization techniques.
The connectivity of individual data objects is further modeled with an ontology and stored in a Knowledge Graph (cf. Section~\ref{sec:ontology}).

MatInf also includes an API wrapper implemented in Python that simplifies the retrieval of stored data without the need to handle raw SQL API calls or the use of the GUI.
Currently, the API supports the retrieval of objects, metadata, and related files; however, an ingest functionality is under development.
The API wrapper is particularly useful for data analysis, as results can be directly integrated into computational workflows.
It also provides a low barrier to involving students for their projects or theses based on data-driven methods by providing easy access to the most recent data from the consortium, thereby allowing them to contribute to cutting-edge research.

Finally, MatInf also allows for granular user control.
Access to the GUI for data ingest or download is only possible for registered and logged-in users.
An application for registration requires approval by an administrator.
API access is handled via a user-specific API tokens.

\section[Ontologies and Knowledge Graphs]{A semantic layer for the RDMS: Ontologies and Knowledge Graphs}
\label{sec:ontology}
To increase the interoperability and lower the barrier of entry to the experimental data and provenance metadata contained within MatInf, we developed a parallel ontological representation.
Ontologies, formally defined by Studer et al.~\cite{ontologyrudi} as \emph{``a formal, explicit specification of a shared conceptualization''}, are composed of the following elements:
\begin{itemize}
    \item \emph{Individuals}, consisting of uniquely identifiable entities, e.g. \tcbox[entity]{EDX measurement 1458} or \tcbox[entity]{A06 Project}.

    \item \emph{Classes} to which individuals belong to, enabling their categorization. Following the example above, \tcbox[entity]{EDX measurement 1458} would belong to the \tcbox[crc]{Measurement} class, and \tcbox[entity]{A06 Project} would belong to the \tcbox[crc]{Project} class.

    \item A set of all the possible \emph{relations} between individuals that we can employ. For example, we can use the relation \tcbox[pmd]{was performed by} to express that \tcbox[entity]{EDX measurement 1458} \tcbox[pmd]{was performed by} \tcbox[entity]{A06 Project}.

    \item A collection of logical axioms that further define and restrict the classes and relationships between individuals. In this case, we can define that \tcbox[crc]{Project} is a subclass of a wider \tcbox[crc]{Organization} class.

    \item Formal semantics that enable machines to reason based on these axioms, and thus surface new relationships. For example, given the axiom defined above, we can infer that, since \tcbox[entity]{A06 Project} is a \tcbox[crc]{Project}, then it is also an \tcbox[crc]{Organization}.
\end{itemize}

This simple but flexible representation of data allows us to establish reusable, semantically-rich and machine-readable vocabularies for any domain.
In more practical terms, the research data produced by different institutions can be made easily interoperable if both of them express it with the same ontologies.  For these reasons, the use of ontologies and Knowledge Graphs as part of research data management is becoming increasingly popular to drive the FAIRification of heterogeneous research data~\cite{bensmann2020infrastructure}, particularly in the interoperability criteria. Moreover, we can further represent this data within a unique, integrated collection by employing a \emph{Knowledge Graph} that follows these ontologies as its data schema.  
In materials science, we can already find works that employ ontologies and Knowledge Graphs for these purposes.  Hernandez et al.~\cite{hernandezDataIntegrationFramework2024} employ ontologies to express Laser-powder bed fusion (L-PBF) data as machine-readable JSON-LD files. Similarly, Bayerlein et al.~\cite{bayerleinSemanticIntegrationDiverse2024} construct a Knowledge Graph that aligns and unifies data from tensile tests and darkfield transmission electron microscopy, revealing correlations between mechanical and microstructural properties in the process.
Moreover, ontologies can allow data integration from electronic lab notebooks (ELNs), as shown in the mechanical testing~\cite{schillingSeamlessScienceLifting2025} and flame spray pyrolysis~\cite{vollbrechtIntegratedDataPipeline2025} fields.
At the same time, ontologies are proving to be the key for a similar FAIRification of computational workflows, as shown by projects such as StahlDigital~\cite{rotersStahlDigitalOntologyBasedWorkflows2025}, where pyiron workflows are semantically described and integrated with an existing data source.

\subsection{Developing an ontology for CRC1625}
The ecosystem of ontologies in materials science has grown steadily in the past years, allowing us to reuse and extend established ontologies in the field. Particularly, the Platform MaterialDigital Core Ontology (PMDco)~\cite{PMDco} forms the basis for our ontology, as its goal of modeling both experimental results and workflows aligns with the nature of our data space. This allows us to ontologically represent and enrich not only research objects (i.e., materials libraries and samples) and experimental results, but also the handover workflows contained in MatInf.
Handover workflows form an ordered provenance record of the combined work of individual CRC1625 projects, in which a given materials library or sample has been analyzed or treated. With the aim of further exploiting these provenance records, we use the Knowledge Graph to represent the same sample-centric view of the data in MatInf, and then enrich it by interconnecting it with handover events.
This allows us to conveniently visualize and access experimental data based on (1)~a sample-centric perspective, where we can directly access its related experimental data, (2)~a workflow-centric one, where we can navigate experiment plans of a sample as an ordered sequence of handover events, or (3)~as a combination of both. 

\begin{figure}[htp!]
  \centering
  \begin{minipage}[t]{.49\linewidth}
    \includegraphics[width=\linewidth]{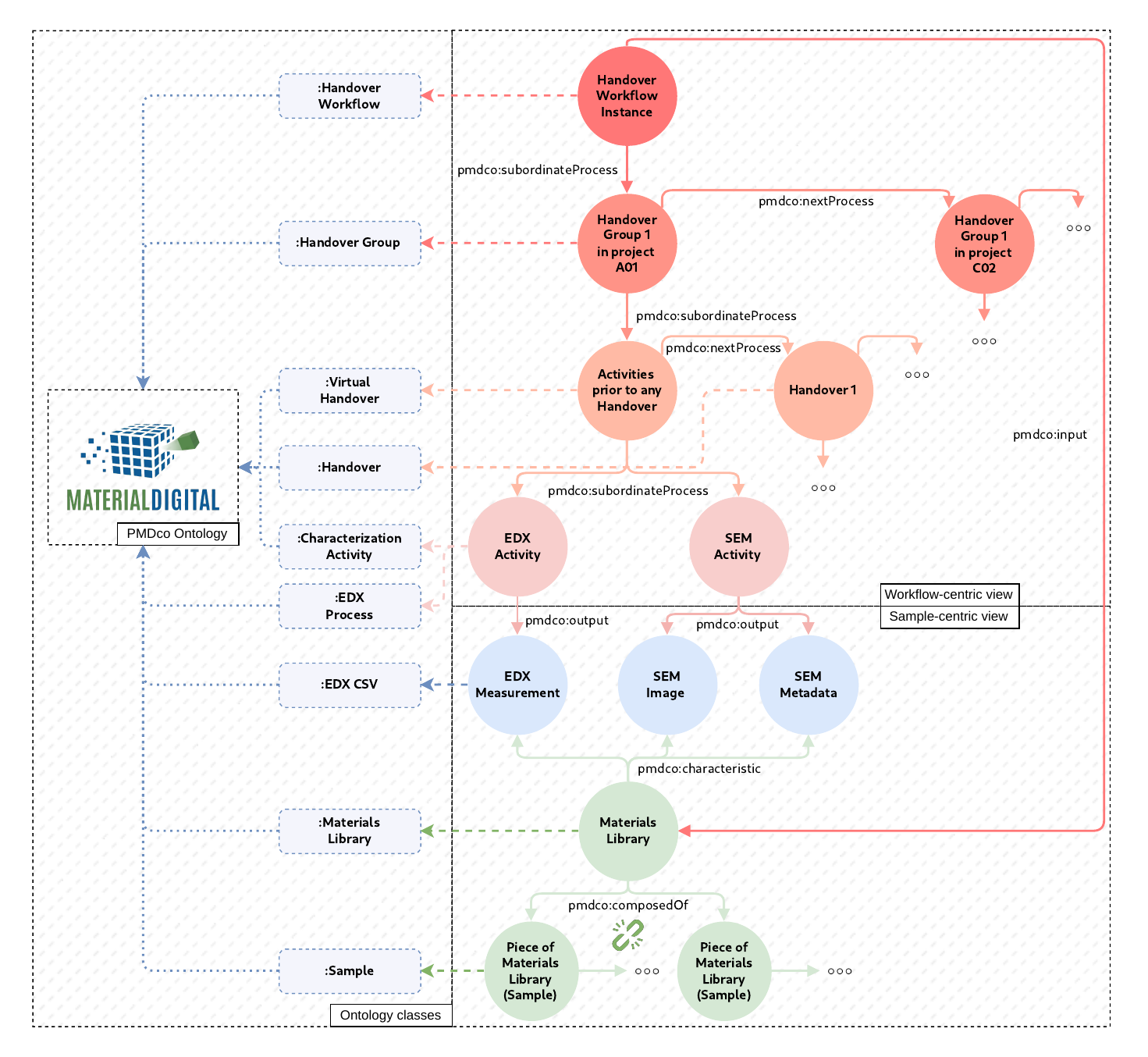}
    \caption{Sample- and workflow-centric views of the data contained in MatInf (bottom and top parts respectively, as individuals), and their correspondence to ontology classes (left). Dashed arrows indicate class memberships, while dotted arrows indicate subclass memberships to PMDco classes as part of their alignment. Note that any other attributes or metadata records are not shown for clarity.}
    \label{fig:ontology_overview}
  \end{minipage}
  \hfill
  \begin{minipage}[t]{.49\linewidth}
        \centering
        \includegraphics[width=0.65\linewidth]{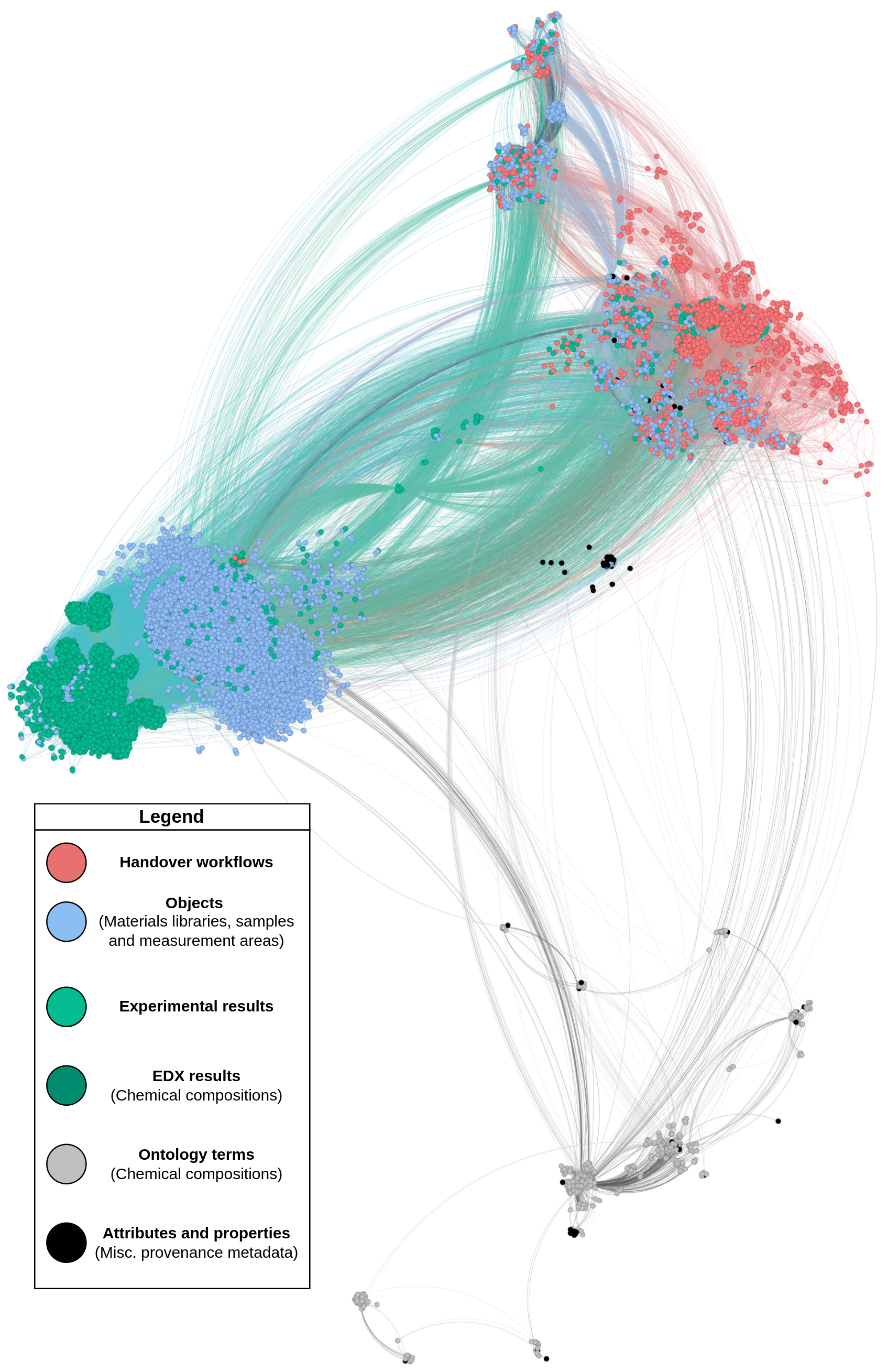}
        \caption{Rendering of the CRC1625 Knowledge Graph. Nodes are clustered together and colored according to their class indicated in the legend. Edges are color-coded according to their source nodes. Note that clusters are highly dense and edges (while they represent semantic connections) are unlabeled in this figure for clarity.}
        \label{fig:kg_viz}
  \end{minipage}
\end{figure}

An example of this representation is shown in Figure~\ref{fig:ontology_overview}. Notably, the handover workflow follows a four-level hierarchy, with each successive level providing a greater level of detail: \emph{handover groups} encompass successive handover events within the same group or project in the CRC1625, while individual handovers contain the different characterization \emph{activities} undertaken, which in turn encompass individual experimental results that belong to the same characterization technique. The ontology additionally represents other provenance records and structures present in MatInf, such as the discretization of materials libraries into MA grids.

\subsection{Knowledge Graph CCSS surfaces}
Following the ontology as its \emph{data schema}, we can create a Knowledge Graph that is instantiated with the data from MatInf.
This is achieved by developing an Extract, Transform and Load (ETL) pipeline that primarily employs declarative \emph{mappings}.
These mappings, defined via the RML language~\cite{rmlspec} and executed through the RMLMapper~\cite{rmlmapper} and RMLStreamer~\cite{rmlstreamer} tools, convert results of queries against its relational database to RDF~\cite{rdfspecs} triples, the standard representation of Knowledge Graphs. A visualization of the resulting Knowledge Graph is shown in Figure~\ref{fig:kg_viz}, highlighting the highly interconnected nature of the research data within MatInf.

The creation of this Knowledge Graph can provide to several additional functionalities to the RDM platform offered to the CRC1625, which are currently being investigated:

\begin{itemize}
    \item The representation of the relational database as a Knowledge Graph with a structured but semantically-rich ontology allows the expression of complex information needs as simple SPARQL\footnote{\href{https://www.w3.org/TR/sparql11-query/}{SPARQL 1.1 specification} (Last accessed: October 2025).} queries. Similar to prior approaches for querying large-scale Knowledge Graphs over heterogeneous infrastructures~\cite{heling2018querying,heling2022federated}, our system exposes a flexible data access layer that abstracts the underlying data sources while retaining semantic expressiveness. In practice, this allows us to define data access APIs through the Knowledge Graph that allow their end users to intuitively modify its underlying SPARQL queries if necessary. An example of such queries is shown as two Listings in Figure~\ref{lst:sparql}. 
    
    \item Aside from traditional data access methods, a graph representation allows hierarchical visualizations of a Knowledge Graph, enabling users to, e.g., navigate experimental workflows.

    \item The use of SHACL~\cite{shacl_spec}, a powerful data validation language for Knowledge Graphs, enables us to define complex restrictions on the research data~\cite{ke2024efficient}. We are currently investigating its use to define and validate ideal experiment plans against the existing handover workflows in MatInf. 
\end{itemize}

\begin{figure}[h]
    \centering
    \begin{minipage}[t]{.49\linewidth}
        \begin{bluebox}{}
            \begin{minted}[fontsize=\scriptsize]{sparql}
# IRI prefixes (usually hidden from the user)
PREFIX pmdco: <https://w3id.org/pmd/co/>
PREFIX : <https://crc1625.mdi.ruhr-uni-bochum.de/>
PREFIX project: <https://crc1625.mdi.ruhr-uni-bochum.de/project/>

SELECT ?handover ?other_activity
WHERE {
  # Handover groups that took place in the A06 project
  ?handover_group a :HandoverGroup.
  ?handover_group prov:wasAssociatedWith project:A06.
  ?handover_group 
    pmdco:subordinateProcess/pmdco:nextProcess* 
    ?handover. # Any handover in the group's handover chain

  # The XRD technique must have been applied in the handover
  ?handover pmdco:subordinateProcess ?activity.
  ?activity a :XRDProcess.

  # And another (non-XRD) technique was also applied
  ?handover pmdco:subordinateProcess ?other_activity.
  FILTER (?activity != ?other_activity)
}
            \end{minted}
        \end{bluebox}
\captionof{listing}{\emph{``Given all handovers that took place in the A06 project and where XRD was employed, retrieve those where another measurement technique was also used''}.}
    \end{minipage}
    \hfill
    \begin{minipage}[t]{.49\linewidth}
        \begin{bluebox}{}
            \begin{minted}[fontsize=\scriptsize]{sparql}
# IRI prefixes (usually hidden from the user)
PREFIX pmdco: <https://w3id.org/pmd/co/>
PREFIX prov:  <http://www.w3.org/ns/prov#>
PREFIX foaf:  <http://xmlns.com/foaf/0.1/>


SELECT 
    ?user_name 
    (COUNT(?measurement) AS ?number_of_measurements_authored)
WHERE {
  ?measurement a pmdco:ValueObject.

  ?measurement prov:wasAttributedTo ?user.
  ?user foaf:name ?user_name.
} 
GROUP BY ?user_name 
ORDER BY DESC(?number_of_measurements_authored)





            \end{minted}
        \end{bluebox}
\captionof{listing}{\emph{``Return the number of measurements uploaded to MatInf by each person, in descending order''}.}
    \end{minipage}
    \caption{Examples of a complex information needs expressed as SPARQL queries in the Knowledge Graph.}
    \label{lst:sparql}
\end{figure}

Finally, by additionally aligning our ontology with other ontologies such as EMMO~\cite{EMMO}, or embracing semantic data interchange formats, we aim to make the publicly available parts of our research data interoperable with other related datasets and, therefore, reusable at scale.

\section{FAIR assessment}
The FAIR guiding principles were first introduced in 2016 by the FORCE11 group \cite{wilkinson2016fair}. Since then, the need to evaluate and assess the FAIR maturity of digital resources has become increasingly important \cite{wilkinson2019evaluating}. These principles, defined for both data and metadata, are widely adopted across research communities, leading to the development of specific metrics that determine whether a system can be considered FAIR. In this context, we explore several available FAIR assessment tools.  
The assessment tools include FAIRshake \cite{gaignard2023fair}, automated evaluators such as F-UJI \cite{devaraju2021automated}, and manual checklists like the FAIRsFAIR guidelines \cite{fairsfair2021framework}.
Each tool offers different levels of automation, customization, and user interaction.
Given the diverse technical backgrounds within the CRC1625 team, we select the FAIR Data Self-Assessment Tool developed by the Australian Research Data Commons (ARDC) \cite{ardc2020fair}.

This tool is a questionnaire-based resource developed for educational and informational purposes, consisting of a single web page that lists all questions.
It uses a mix of yes/no and multiple-choice single-answer questions divided into four sections, one for each FAIR principle.
Guidance is provided through a pop-up description when clicking on each question.
The nature of the questions and their multiple-choice format helps with the comprehension of the assessment process.
Such a critical self-assessment helps us identify both strengths and areas where improvements are needed.

Filling out the questionnaire to the best of our knowledge led to a score of 77\,\%\footnote{\href{https://ardc.edu.au/fair-tool-report?token=62ac9330affa5c1ea881df7012509ae1}{Link to self assessment results, October 27th, 2025.}}.
The main reasons for a not-higher score is that most of our data is only available for authorized users and we currently do not have a default permissive license associated with each data record.
In addition, our metadata formats are currently not standardized at scale which, however, will change once we include the Knowledge Graph (c.f. Section~\ref{sec:ontology}) on top of the relational database which links our data with existing data models.
An assessment like this supports internal discussions on how to better structure and document our data, adopt shared standards, and improve long-term reusability.
Therefore, the assessment is not just a way to measure FAIRness but a valuable tool for guiding our next steps in building a more robust and collaborative data infrastructure.

\section{Usability and user training}
In the CRC1625, we built a reliable RDMS but also strive to make it intuitive and easy to use.
Our approach is inspired by human-computer interaction and user experience design, meaning that researchers directly shape how the system develops.
By regularly sitting down with users and running usability tests, we get a clear picture of how the RDMS works in real research settings.
User feedback shows us what is working, what is frustrating, and where we can make things smoother so the system keeps evolving in ways that support their day-to-day work.
We also align our work with national and international efforts such as NFDI-MatWerk and the broader FAIR data community, helping to strengthen awareness and encourage the adoption of good data management practices (``interoperability'').
By combining usability testing with user training, CRC1625 ensures that the RDMS grows in step with its users supporting both everyday research work and the long-term goals of FAIR and sustainable data stewardship.

\subsection{Usability}
Usability is a key concept in human-computer interaction (HCI), referring to the degree to which a system enables users to achieve their goals effectively, efficiently, and with satisfaction~\cite{bevan2015iso}.
In the context of an RDMS, (good) usability is essential to motivate researchers to participate by reducing the time and effort required from individuals and to enable an intuitive interaction with the platform.
Usability testing evaluates whether the product can be used easily by target users~\cite{nielsen1994usability}.
In the context of software interfaces, usability is determined by considering target users’ characteristics and the specific tasks required by the interface.
Therefore, designing highly usable software applications requires an in-depth understanding of target users and their tasks \cite{darejeh2024critical}.
Cognitive load measurement methods typically serve as an efficient technique for conducting accurate usability evaluations.
Cognitive load is evaluated using subjective tools such as the NASA Task Load Index (NASA-TLX)~\cite{hart1988development} and the Paas Mental Effort Rating Scale~\cite{paas1992training}.
Objective performance metrics such as task completion time and error rates, and physiological measurements such as eye tracking and heart rate variability complement the subjective assessment.
All these methods help to determine how mentally demanding a task or system interface is for users.

Within CRC1625, we are currently planning a usability test conducted through online meetings.
Prior to the usability sessions, participants receive a short training that introduces the FAIR data principles, the key functionalities of the RDMS, and the object-based structure of data creation within the system.
This orientation ensures that all users have a baseline understanding of both the theoretical and practical aspects of research data management.
During the usability sessions, ten representative users perform a set of predefined typical tasks using the RDMS platform.
The usability sessions are recorded to enable post-analysis of user interactions, including task completion time, error frequency, and navigation behavior.
In addition to performance-based metrics, participants complete the NASA Task Load Index (NASA-TLX) questionnaire to evaluate their perceived cognitive workload.

The results of this usability test will provide both objective and subjective insights into the strengths and weaknesses of the current RDMS GUI implementation.
Furthermore, the findings will help to identify gaps in users’ understanding and highlight areas where additional training or interface improvements are needed to improve adoption and the overall efficiency of research data management processes.

\subsection{User training}
User training ensures that researchers understand both the technical functionalities of the system and the underlying principles of good data management.
Well-designed training programs not only enhance users’ ability to take full advantage of RDMS features but also increase confidence, reduce errors, and promote long-term adoption.
Together, usability and user training play a key role in enabling researchers to manage, share, and reuse data more effectively.
Our training focuses on three main areas: (1) the practical operation of the RDMS interface, including data upload, linking, and metadata management; (2) the conceptual understanding of FAIR principles and their application in daily research workflows; and (3) advanced functionalities such as version control, data citation, and interoperability with analysis or visualization tools. 

Given that CRC1625 involves researchers from diverse disciplinary backgrounds, training activities are designed to establish a common baseline of knowledge and competencies across all users.
This shared understanding helps bridge differences in technical experience and domain expertise, allowing participants to collaborate more effectively and contribute equally to the research data management process.
Our training formats include interactive workshops, guided tutorials, self-paced online modules, and one-to-one sessions.
Continuous refresher sessions during retreats and feedback-based updates to the training materials maintain high usability standards and support the sustainable adoption of the system across the CRC1625 consortium.
To ensure inclusivity, training materials come in various different formats: text, slides, and videos.

Additionally, peer-learning sessions and informal `data clinics' complement formal training, enabling experienced users to help others troubleshoot and share practical insights.
Onboarding integration ensures that new CRC1625 members are introduced to data management practices from the beginning.
We are also discussing awarding certificates upon completion of trainings, helping them demonstrate competence in FAIR-aligned data handling beyond the context of CRC1625.
CRC1625’s training strategy aligns with the NFDI-MatWerk initiative and the EOSC FAIR Training framework.
In this context, some members of CRC1625 take part in NFDI-MatWerk’s Task Area ``Workflows, Software, and Demonstrators (TA WSD)'', helping to develop practical FAIR workflows and share best practices with the research community.
In the application for the next funding phase of NFDI-MatWerk, CRC1625 is a participating project.
Our members also participate in community training events such as the 
\href{https://www.eusmat.net/research/other-events/nfdi-spring-school-2024/}{NFDI–MatWerk Spring School 2024 at Saarland University}, 
where they exchange experiences and contribute to strengthening the culture of FAIR data management in materials science.

\section[AI-readiness]{AI-readiness of data and reproducibility}
\label{sec:ai_readiness}
Modern research in materials science increasingly relies on data-driven approaches to accelerate materials discovery, but this is only possible when the underlying data is structured, accessible, and reproducible.
Our goal is to ensure that the diverse, high-dimensional, and often sparse data generated across experiments becomes not only machine-readable but also AI-ready.
In this section we present some of our use cases for the data in the RDMS.

\subsection{From heterogeneous data to unified datasets}
Within CRC1625 we face a common challenge: making sense of the large amounts of heterogeneous, highly correlated data generated across various experiments which is typically sparse.
The RDMS contains over 100 defined data types, yet data exist for only a subset of them, and even those are from a comparatively small number of examples of the order of $100$s of samples from diverse sources with high dimensionality that are sparse, include noise and uncertainty with complex correlations, and are also referred to as \textit{Tall Data}~\cite{Grigo2019}; sometimes Tall Data also implies `unstructured'~\cite{Jablonka2020} which can also be applied to our data space.
For instance, we manage structured numerical data such as chemical compositions and electrical resistance alongside unstructured data like high-resolution SECCM scans and micrographs.
The heterogeneity of data provides a challenge for obtaining a comprehensive view of each individual sample across multiple modalities.

Beyond the technical integration of data, the reuse of data within a collaborative research environment introduces conceptual and ethical challenges.
Data-driven approaches that automatically leverage all available datasets differ fundamentally from interpersonal collaborations, where researchers directly negotiate access, interpretation, and credit.
While the automation aspect of data management improves scalability and practical use, it must also respect the rights and intentions of individual data producers.
To ensure this balance, we are currently implementing a method for automated retrieval of consent and licensing of data and metadata by linking each dataset to its associated terms of use.
Such a mechanism will allow researchers to conveniently interpret which data may be included in unified datasets according to their license-derived permissions or pending approval status.
Some portion of the data are already openly available under licenses such as CC-BY-SA 4.0, while others remain unpublished or restricted until validation or formal publication.
If data is already published, our RDMS is capable of incorporating this information, e.g., through the addition of digital object identifiers (DOIs) as metadata of data objects.
In this way, the RDMS mediates between the principles of open data reuse and the courtesy toward data producers that underpins scientific collaboration and good scientific practice.

Apart from licensing considerations, we implemented a user-driven, API-based pipeline designed to provide AI-ready, structured data with minimal post-processing.
Data retrieval via API begins with authenticated API access, enabling users to selectively download data types that actually contain entries.
Each dataset is constructed by automatically fetching files via their unique `ObjectId' and merging them into a single table.
Depending on the research question, the pipeline can then be extended through a sequence of user-defined steps to improve data quality and usability:

\begin{itemize}
    \item Handling missing values: the pipeline automatically removes empty columns and offers several strategies for missing data treatment, including k-Nearest Neighbors (KNN), interpolation, or statistical measures such as the mean, max, or median.
    \item Dimensionality reduction: users can apply advanced techniques such as Principal Component Analysis (PCA), t-distributed Stochastic Neighbor Embedding (t-SNE), Uniform Manifold Approximation and Projection (UMAP), or autoencoders. Normalization options ensure consistency across features.
    \item Image preprocessing and feature extraction: since image data cannot be directly analyzed numerically, a dedicated routine identifies and crops the region of interest (typically the materials library), resizes images, and converts them into low-dimensional numerical vectors using a pre-trained autoencoder. These vectors can then be integrated with other numerical features such as composition and resistance.
\end{itemize}

Naturally, the choice of analysis methods depends on the research question, but through this modular design, users can build AI-ready datasets in a reproducible and transparent manner, significantly lowering the barrier to applying machine learning to complex experimental data because it all starts with a documented retrieval of data.
Such uses of data are also very much aligned with the concept of reproducible workflows~\cite{Janssen2025}.

\subsection[Mapping materials data space]{Mapping the materials data space and ensuring reproducibility}
\begin{figure}[htp!]
\centering
\begin{subfigure}{.32\textwidth}
    \includegraphics[width=\textwidth]{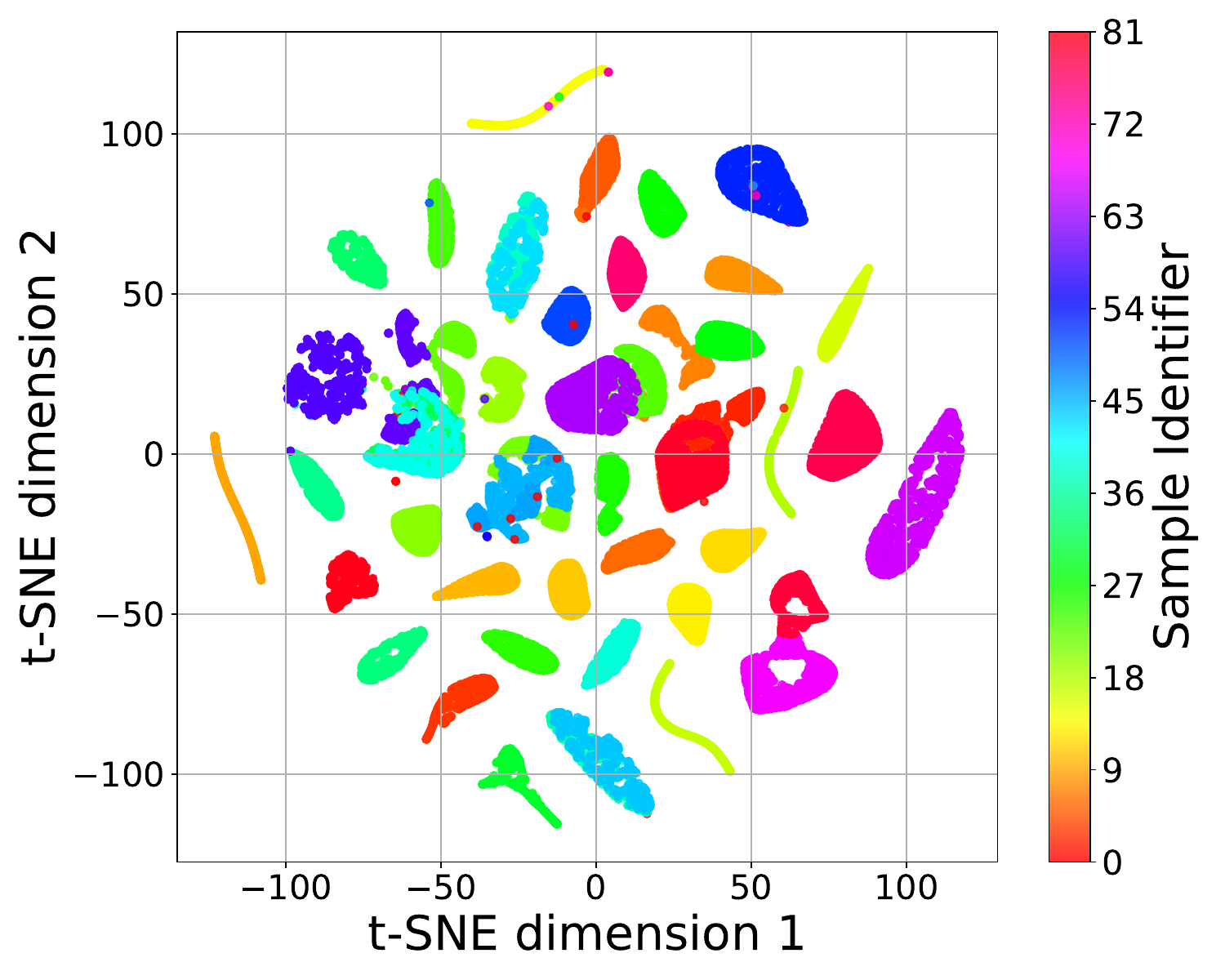}
    \caption{Sample IDs.}
    \label{fig:tsne_sampleid}
\end{subfigure}
\hfill
\begin{subfigure}{.32\textwidth}
    \includegraphics[width=\textwidth]{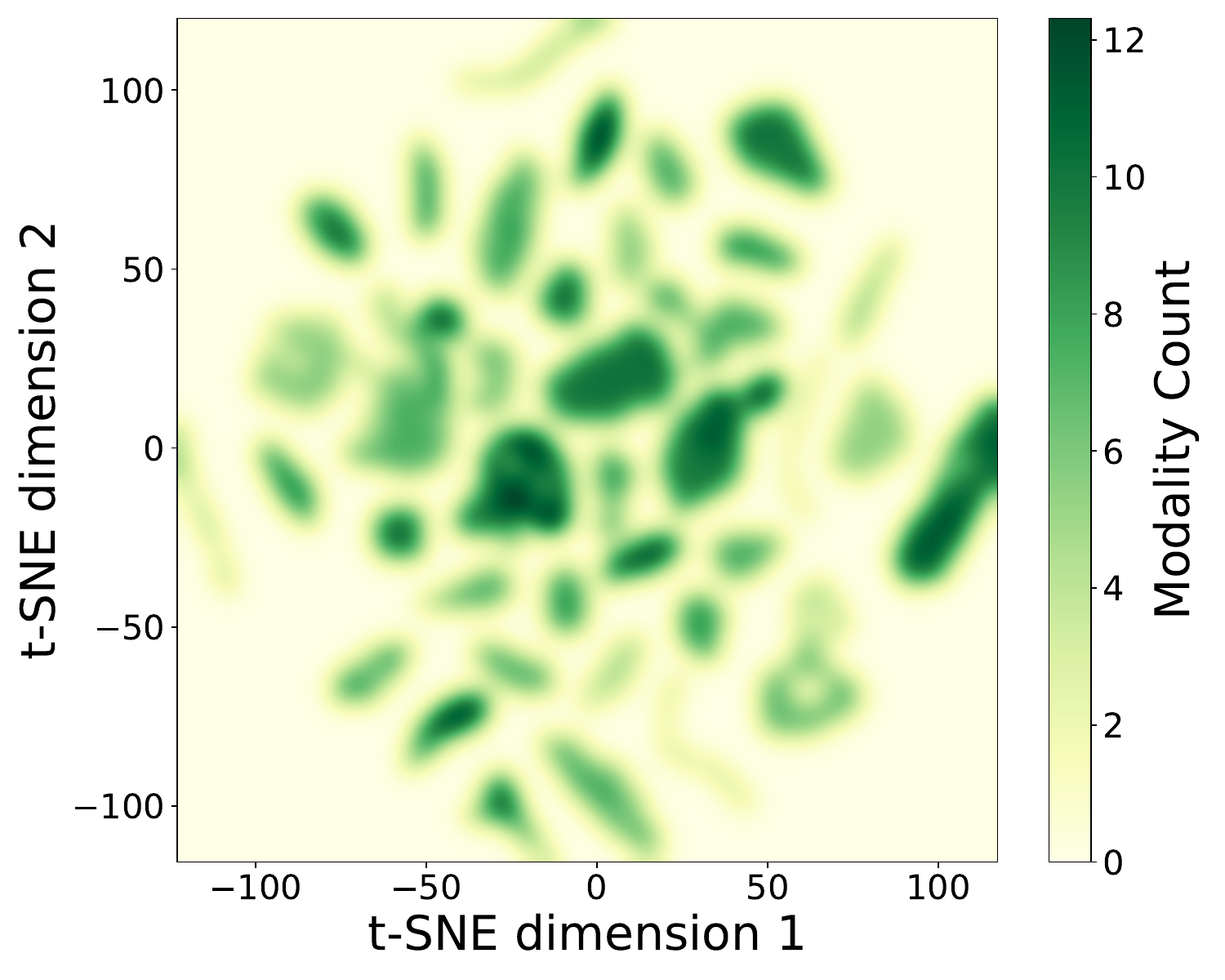}
    \caption{Number of modalities.}
    \label{fig:tsne_no_of_modalities}
\end{subfigure}
\hfill
\begin{subfigure}{.32\textwidth}
    \includegraphics[width=\textwidth]{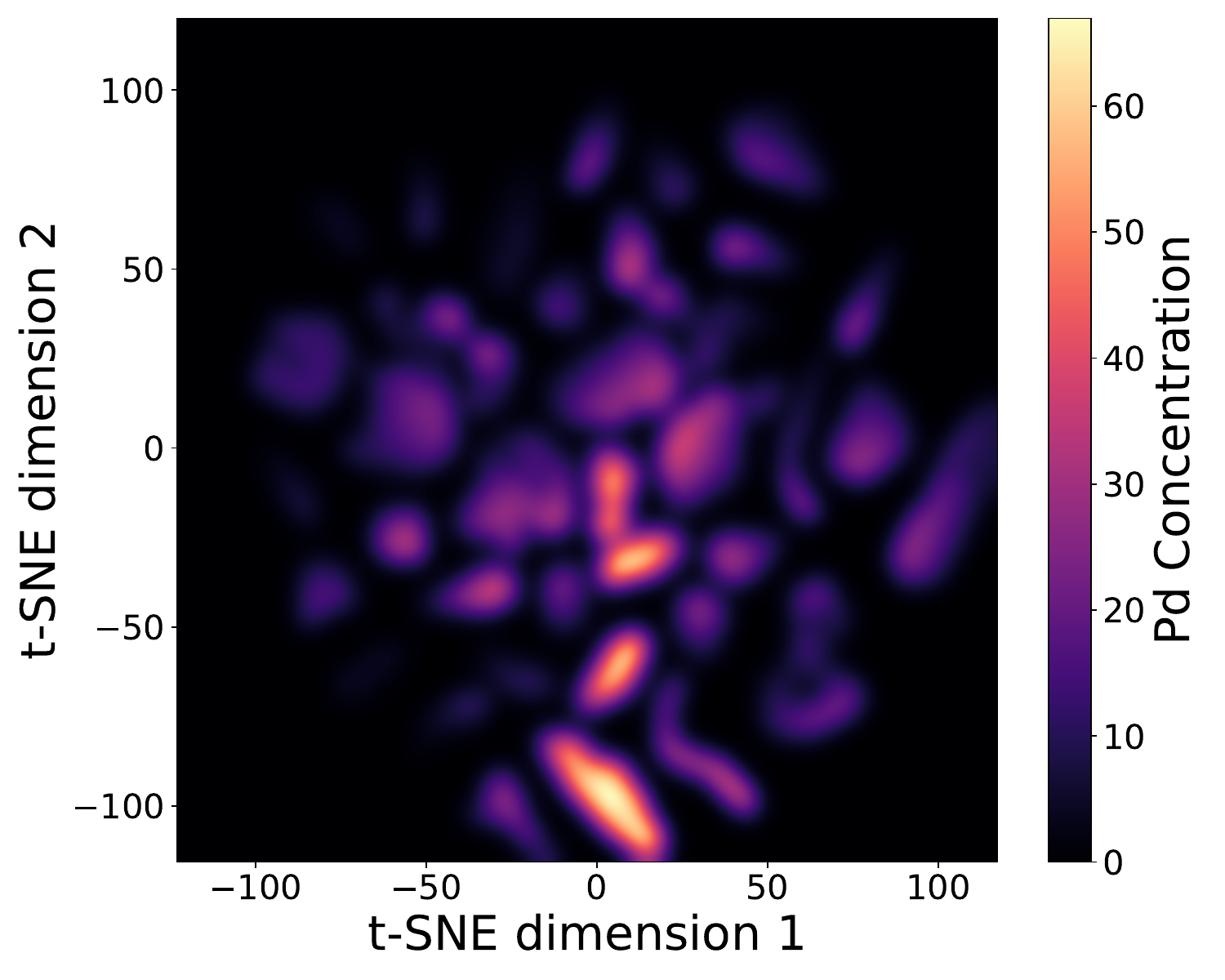}
    \caption{Pd content.}
    \label{fig:tsne_pd_content}
\end{subfigure}
\caption{Two-dimensional maps of the 13-dimensional chemical space (cf. Figure~\ref{fig:matinf_gui_periodic_table}, non-faint elements) using t-SNE with different color schemes where each data point corresponds to one of 19888 MA: Subfigure~\ref{fig:tsne_sampleid} shows a coloring of the sample IDs, subfigure~\ref{fig:tsne_no_of_modalities} a heatmap how many different data modalities exist for a specific MA, and subfigure~\ref{fig:tsne_pd_content} shows a heatmap of the Pd content corresponding to the most prevalent element in our materials libraries.}
\label{fig:data_landscape_dimensions}
\end{figure}

Besides mapping the relationship within our data space through a Knowledge Graph as in Figure~\ref{fig:kg_viz}, we can also visualize the chemical space covered by our data.
By focusing on individual MAs as an entity (cf. Figure~\ref{fig:wafer_data_dimensions}), we can create a 13-dimensional composition vector referring to the concentrations of Ag, Au, Pd, Pt, Rh, Ru, Cu, Ir, Ni, Mg, H, Ti, and Re for each of the 19888 MAs.
Using t-SNE to reduce the dimensions to two allows us to map the chemical space as shown in Figure~\ref{fig:data_landscape_dimensions} where all subfigures use the same two t-SNE dimensions.
This is in line with one of the hypothesis of the CRC1625: The main tuning parameters to affect surface properties is the compositions because we assumes homogeneous FCC crystal structure across all samples.
We can then overlay other modalities across the existing materials data space, such as the Sample ID (Subfigure~\ref{fig:tsne_sampleid}), an interpolated view of the number of different data modalities existing for each MA (Subfigure~\ref{fig:tsne_no_of_modalities}) showing the level of sparsity and data heterogeneity, or a very specific property of each MA, the Pd content (Subfigure~\ref{fig:tsne_pd_content}).

Such datasets enable flexible AI-driven studies, for example, predicting electrical resistance from chemical composition~\cite{Stricker2022e} or classifying synthesis conditions based on micrograph- or photograph-derived features.
By ensuring that each step in the data preparation process from acquisition to preprocessing is transparent and reproducible, our RDMS directly contributes to the objectives of NFDI MatWerk: transforming dispersed, highly correlated, sparse, high-dimensional raw data into structured, reusable resources that accelerate discoveries in materials science.

\subsection{Text mining for composition-property prediction}
The (very) high-dimensional chemical search space of possible candidate catalyst compositions is prohibitive even for the fastest high-throughput methods.
Consequently, strategies for directed search and the prediction of composition-property relationships with minimal data are required.
One of our employed solutions is based on text mining of latent knowledge~\cite{Zhang2024} and uses composition representations based on word embeddings in conjunction with experimental data~\cite{Zhang2025c} to extrapolate composition-property trends from ternary to quaternary materials systems.
The experimental data is, of course, taken from the RDMS and used in a reproducible workflow.
In addition, we (re-)use available experimental data to fine-tune text mining approaches based on multi-objective optimization which require no experimental input to identify reaction-specific high-performing candidate compositions~\cite{Zhang2025a}.
This reduces a given candidate pool by one to two orders of magnitude, and we can directly request a synthesis of a new sample through the RDMS.

\subsection{Active learning}
\label{ssec:activel_learning}
\begin{figure}[htp!]
    \centering
    \includegraphics[width=.75\textwidth]{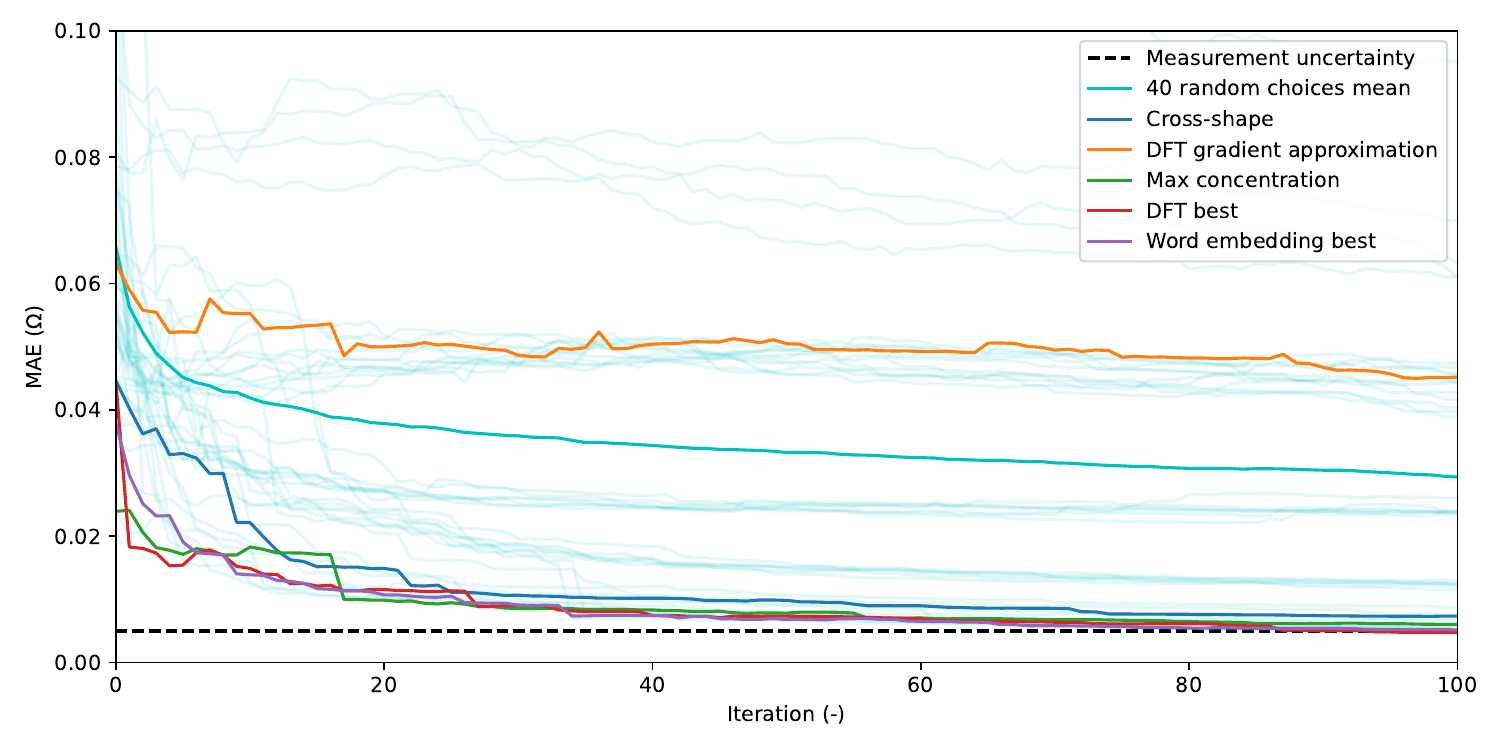}
    \caption{Mean absolute error (MAE) evolution for active learning loops with different cold starts (initializations). Black dashed line indicates the measurement uncertainty. Please refer to the text for further details; reproduced from~\cite{Stricker2022e}, licensed under \href{https://creativecommons.org/licenses/by/4.0/}{CC-BY 4.0}.}
    \label{fig:mae_vs_iteration_active_learning}
\end{figure}

Once a promising range of candidate compositions in a materials system is identified, the measurement time for a specific property on a specific materials library can be optimized by use of active learning~\cite{Thelen2023,Thelen2025b}.
This comes at the expense of not measuring everything but measuring the most useful data points for a model to predict the rest with an acceptable uncertainty.
The initialization of active learning loops, commonly known as the cold start problem~\cite{Maltz1995,Schein2002}, can further be optimized~\cite{Stricker2022e} through prior knowledge.
Figure~\ref{fig:mae_vs_iteration_active_learning} demonstrates this acceleration based on different cold start strategies: Compared to a baseline of 40 (faint cyan) averaged (dark cyan) random cold starts and the previous best initialization strategy of a cross shape (blue), our strategic initializations based on the maxima of elemental concentrations and existing correlations based on DFT calculations and text-derived latent knowledge, the active learning loop for the 342 MAs' resistance is accelerated 7-fold.
In other words, instead of measuring 342 MAs individually, only approx. 50 are measured, and the rest are interpolated.

\begin{figure}[htp!]
    \centering
    \includegraphics[width=0.8\textwidth]{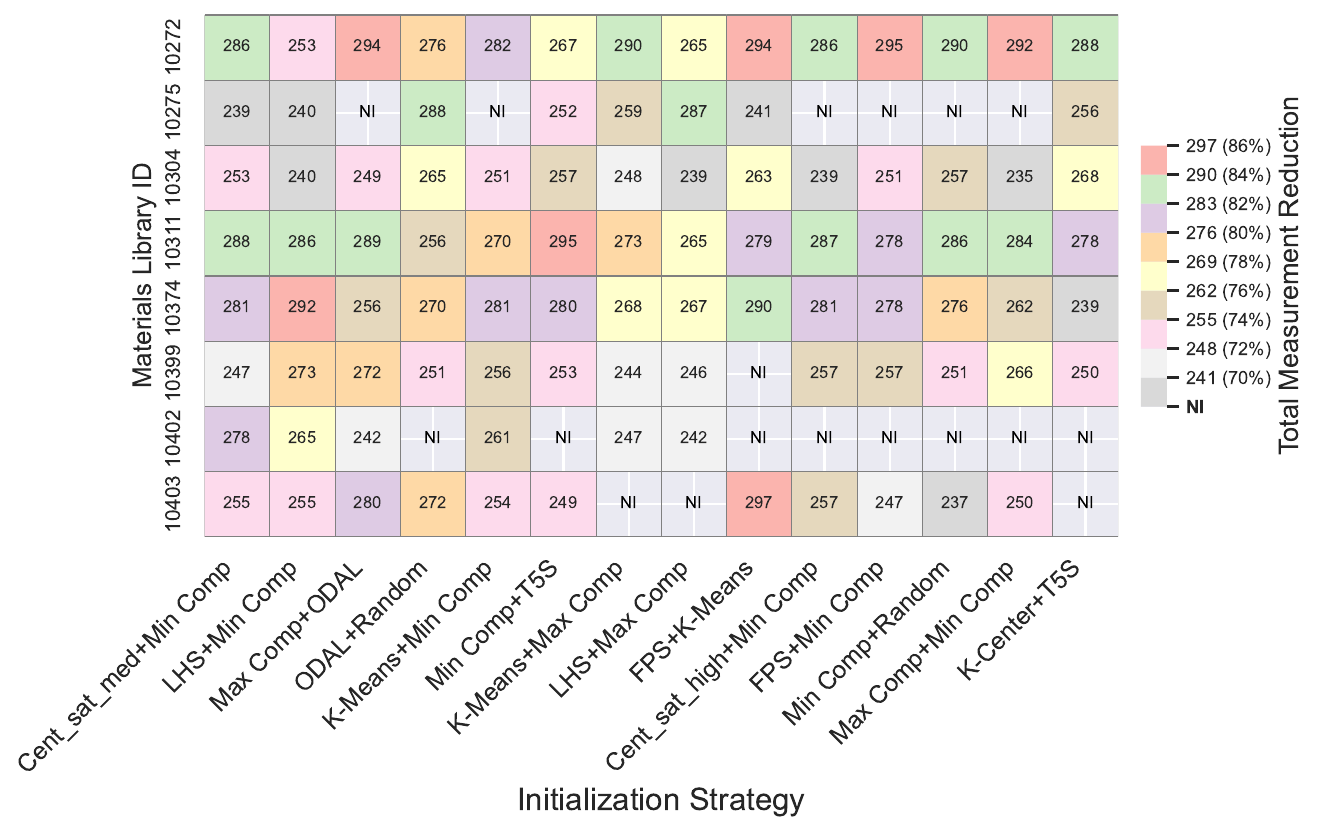}
    \caption{Overview of the reduction of the number of measurements for different initialization strategies that reduce the measurements needed by 85\% down to 60\% across materials libraries with Uncertainty Sampling (maximum 342 per materials library).  Abbreviations: Cent\_sat\_high – high saturation centroids; Cent\_sat\_med – medium saturation centroids; Cent\_sat\_low – low saturation centroids; T5S – Top-5 similarity; Random – Random sampling; Min Comp – Minimum composition; Max Comp – Maximum composition; ODAL – Outlier-based sampling; FPS – Farthest Point Sampling; LHS – Latin Hypercube Sampling; NI – No improvement, which indicates no convergence within the 100 iterations.}
    \label{fig:heatmap_active_learning}
\end{figure}

We are currently in the process of strategically assessing different strategies using more than one existing data modality and testing which combinations work best to cold start active learning loops for expensive and/or time-consuming measurements.
The result of this survey will subsequently inform which (cheaper and/or faster) characterization techniques should preferentially be employed because of their downstream usefulness, further optimizing the use of available resources and time.
Figure \ref{fig:heatmap_active_learning} summarizes the reduction in measurements achieved by various multimodal initialization strategies.
Some approaches can reduce the required number of measurements from 85\% down to 60\% across different materials libraries.

These examples represent a selection of data-driven approaches we are currently exploring to tackle the high-dimensional problem space of the atomic-scale understanding and design of multifunctional compositionally complex solid solutions surfaces with complex correlations given limited resources.

\section{Conclusions and future}
CRC1625 started in April 2024.
The current state of our RDMS benefits from prior developments in individual groups (e.g.,~\cite{Dudarev2025,Banko2020}), and our considerable effort since the start has made it much more useful, in particular the addition of an API, its Python-based wrapper, and the ontology and Knowledge Graph developments.
The \textit{Key Lessons Learned} can be summarized as

\begin{itemize}
    \item A sample-centric data hierarchy enables cross-institutional and cross-modal linking. This mimics the physical world, mainly from the view of experimentalists, and is logical for users. At the same time we can extract data-centric views (cf. Figure~\ref{fig:data_landscape_dimensions}) for data-driven workflows across samples, MAs, and data modalities.
    \item Data creator/authorship as explicit metadata allows relatively simple crediting across a large consortium without the need for one-one or one-many coordination efforts, particularly for the projects using the data. At its core, this is a socio-technical challenge: \emph{``If I share my valuable and expensive-to-obtain experimental data with a consortium, will I receive appropriate credit for my contribution?''} which has to be solved.
    \item AI-readiness is not just the data and its connectivity but convenient API access through a wrapper.
    \item Community-wide developments around RDM, such as the PMDco or EMMO ontologies or the definition of best practices~\cite{AvilaCalderon2025} with examples, are very useful; their adaption and practical implementation in a specific setting, however, require a substantial effort.
\end{itemize}

However, as characterization methods, data models, machine learning and AI methods evolve, a continuously useful RDMS will require constant development and adaption.
Our individual research projects are dedicated to \emph{proper} RDM, and we are currently in a position as a consortium to address those changes.
But one thing becomes increasingly clear: RDM as a task for a consortium or even a whole community hugely benefits from a constant core group of people dedicated to its administration and maintenance, which does not align with fixed, even longer-term funding phases like a CRC.
To support the long-term sustainability and useful impact of our CRC1625 RDMS, several development goals are currently planned.
These future efforts focus on improving usability, expanding automation, and deepening the integration of data analytics, with a strong commitment to reproducibility and FAIR principles:

\begin{itemize}
    \item User experience and interface improvements: We are currently developing a more user-friendly and interactive GUI to further simplify the process for researchers to upload their experimental data. This is particularly important for comparatively low-volume but very expensive and useful experimental data.
     \item Efficient ingest of more data and more data modalities: From Table~\ref{tab:char_techniques} it is clear that even though we mainly operate in a high-throughput approach, the volume of data ingested is manageable. Managing the results from high-throughput characterization was and still is our main concern simply based on the fact that any improvement for these kinds of data has the highest impact on the number of data points. However, the integration of very high-resolution techniques (EC-STM, STM, TEM) will be addressed, too.
    \item Development of representations for high-dimensional materials correlations: One, if not the biggest data science challenge in CRC1625 is to combine all the different data modalities within CRC1625 to provide guidance for experiments and derive design rules for CCSS in electrocatalytic applications. For this, each modality needs to be converted from its raw form (e.g., a micrograph or a spectrum) into a representation amenable for machine learning purposes. A \textit{good} representation of different modalities allows them to construct high-dimensional feature vectors for each composition, which in turn allows the navigation and exploitation of the compositions-property space.

    \item API access improvement: For internal use but also future public access to our data, we are currently collecting experience with reproducible workflows and practical use cases. This experience informs the future development of our data. It is particularly challenging to integrate different data modalities (image formats, spectra, numerical values, etc.). One of our approaches to address this is the development of contrastive learning techniques to align different data modalities~\cite{Mirza2025}.
    
    \item External services designed to provide in-depth support for research data types and meeting the RDMS specification can be seamlessly integrated with the system. These services enable document validation and data import into RDMS, as well as integration with external web applications for data visualization and analysis.

    \item Live instrument integration: We are currently exploring ways to directly connect measurement instruments for real-time data ingestion, automatic metadata capture, and streamlined sample tracking. Other applications are the sample-specific, optimized cold starts for active learning (cf. Section~\ref{ssec:activel_learning}). Concretely, this means an active learning algorithm running on local instrument hardware can query the RDMS via the API for an optimal initialization given the sample under consideration.

    \item Integration of the Knowledge Graph within the RDMS: Our efforts in the development of the semantic layer for our data and the subsequent construction of the Knowledge Graph will in the near future change how we interact with data, e.g., through SPARQL queries, and how we monitor the research and data recording process by comparing \textit{ideal} experimental workflows with the actual reality.

    \item Publication of reference data sets and reference workflows to reuse data: Apart from data publications along with scientific articles, we are also planning to provide reference data sets in line with the proposed `framework for reference data in materials science and engineering'~\cite{AvilaCalderon2025} including data generation and documentation aspects, data formats/ontologies, and a data description article.
\end{itemize}

We have presented our approach to research data management in the context of a complex collaborative research effort.
Our current solution consists of a relational database with flexibility to accommodate different data formats and types, which allows us to connect data points with locations on actual samples.
On top of this database, we have created a semantic layer (ontology) and a Knowledge Graph which will soon be connected to the RMDS.
We also provided the first examples of how our effort bears fruits for data-driven approaches by showing examples of how the data was used to validate word embedding-based property trends, accelerated characterization through active learning with sample-specific cold start, and a glimpse of how a less sparse data space in the context of CCSS can look.
This field report is a record of the current state and our ideas of how \textit{useful} research data management can look and which improvements we see in our approach.
We look forward to inspiring discussions with other researchers in similar circumstances.

\medskip
\textbf{Acknowledgements} \par
All authors gratefully acknowledge funding by the Deutsche Forschungsgemeinschaft (DFG, German Research Foundation) for CRC1625 -- A01, A05, A06, INF, project number 506711657.
Further, all CRC1625 projects contributing data to the RDMS are acknowledged.
VD and AL acknowledge funding by the DFG for TRR 247 -- INF, project number 388390466.

\printbibliography

\end{document}